# Towards Gaussian processes modelling to study the late effects of radiotherapy in children and young adults with brain tumours.

Angela Davey[1], Arthur Leroy[2], Eliana Vasquez Osorio[1], Kate Vaughan[1], Peter Clayton[3,4], Marcel van Herk[1], Mauricio A Alvarez[2], Martin McCabe[1,4] and Marianne Aznar[1]

[1]Division of Cancer Sciences, Faculty of Biology, Medicine and Health, The University of Manchester, Manchester, United Kingdom.
[2]Department of Computer Sciences, Faculty of Science and Engineering, The University of Manchester, Manchester, United Kingdom.
[3]Division of Developmental Biology and Medicine, The University of Manchester, Manchester, United Kingdom.
[4]The Christie NHS Foundation Trust, Manchester, United Kingdom.

**Abstract**: Survivors of childhood cancer need life-long monitoring for side-effects from radiotherapy. However, longitudinal data from routine monitoring is often infrequently and irregularly sampled, and subject to inaccuracies. Due to this, measurements are often studied in isolation, or simple relationships (e.g., linear) are used to impute missing timepoints. In this study, we investigated the potential role of Gaussian Processes (GP) modelling to make population-based and individual predictions, using insulin-like growth factor 1 (IGF-1) measurements as a test case. With training data of 23 patients with a median (range) of 4 (1-16) timepoints we identified a trend within the range of literature reported values. In addition, with 8 test cases, individual predictions were made with an average root mean squared error of 31.9 (10.1 – 62.3) ng/ml and 27.4 (0.02 – 66.1) ng/ml for two approaches. GP modelling may overcome limitations of routine longitudinal data and facilitate analysis of late effects of radiotherapy.

## 1 Introduction

Each year an estimated 400,000 children are diagnosed with cancer worldwide [1]. Radiotherapy plays a main role in curative-intent treatment with more than 75% of children now surviving more than 5 years [2]. Despite modern radiotherapy techniques, brain tumour survivors have the highest incidence of severe chronic health conditions compared to other childhood cancers, in part due to the impact of radiation on the developing brain [3]. There is a need to study the relationship between radiation to brain substructures and late outcomes so we can plan 'gentler' radiation treatments and ultimately improve patient outcomes.

A common outcome post-radiotherapy in survivors of childhood brain tumours is growth hormone deficiency (GHD), which, if it goes untreated, can lead to short-stature, fatigue and increased risk of diabetes and heart disease among others [4,5]. Routine blood tests are used as follow-up for brain tumour survivors, which include a specific test for Insulin-like growth factor 1 (IGF-1), a primary index for diagnosis of GHD.

To enable the study of the association between dose to brain substructures and GHD, we have so far dichotomised the outcome as the likelihood a child is placed on treatment for GHD (yes vs no). A dichotomous outcome could lead to loss of valuable information on patterns of change for those within each group. In future, we plan to make use of longitudinal measurements of IGF-1 for a more granular study of outcome. Follow-up routine blood tests are typically recorded for children many years post-treatment. However, these tests are infrequent, sampled at irregular timepoints, and subject to inaccuracies. To move towards dose-response modelling with longitudinal data, we need to understand population trends with age and make predictions of an individual's trajectory to provide outcomes at interpolated standardised time-points such that patients can be compared. Gaussian processes (GP) modelling has the potential to enable this.

GP modelling is a machine learning approach that can be used to capture the function that maps any input variable (i.e., time) to any output variable (i.e., IGF-1) [6]. Fundamentally, it is non-parametric regression and more flexible in comparison with typical parametric approaches (e.g., linear and polynomial regression) that are somewhat limited in the spectrum of trends they are able to capture. Due to this flexibility, GPs are particularly powerful in situations where there is limited data and/or relationships are complex. GPs are probabilistic in nature and hence leverage prior information during training to provide uncertainty quantification for the output estimates of the model. Multi-task GPs encompass the situation where a GP may be trained on data from multiple individuals. In this case, one can take advantage of a latent mean process, common to all individuals, to share information across them and enhance modelling and prediction accuracy. This is useful for the situation where some individuals do not have many timepoints sampled or have large gaps in recording, as they can 'borrow' information from other individuals. This would allow us to impute data based on population trends. One package for performing this type of prediction is MAGMA (Multi tAsk GPs with common MeAn) developed by Leroy et al. [7]. MAGMA has proved advantageous in multiple simulated and realistic scenarios (e.g., predicting athletes swimming performance and modelling body mass index during childhood). However, the ability to use such techniques in routine follow-up data for childhood cancer survivors has not yet been explored.

Here, we train multi-task GPs with a common mean to measure how IGF-1 changes with age from routine measurements of children and young adults who received radiotherapy for a brain tumour. We will test **1)** the ability to identify underlying population trends, and **2)** the ability to interpolate individual timepoint predictions in unseen data. By performing these tasks, we highlight the potential of using GPs to impute data at a given time, thereby enabling study of the late effects of radiotherapy.

## 2 Materials and Methods
### Data
We collected data from a single institutional archive on blood tests for IGF-1 recorded after radiotherapy for 31 children and young adults with brain tumours. The dataset was divided quasi-randomly to ensure a 75% / 25% split for training and testing (23 training, 8 test). The quasi-random component was to make sure patients with one timepoint were only included in the training set (as there would not be enough points for testing).

### Training
We trained a MAGMA model, which is defined as the sum of a shared mean GP and an individual specific GP. We test two variations model hyperparameters (HPs) in training: **1)** consider *common* mean and covariance HPs to be equal for all individuals, assuming individuals correspond to different trajectories of the same process, and **2)** individual-specific HPs to generate a more flexible model. Model training is performed via likelihood maximisation within an expectation-maximisation algorithm. To account for the influence of the HPs random initialisation, we repeat the training process 25 times per variation and select the GP models with the maximum log-likelihood. We compare the trained population GP to literature reported values.

### Testing
We ensured that each individual the in the test set had at least two timepoints. Per test patient we randomly sample half of the timepoints for prediction, and half for evaluation. Using the prediction points for each patient, we can predict the entire curve of IGF-1 values with associated uncertainty, in particular at the location of the testing points for evaluation purposes. To evaluate the individual GP model, we calculate the root mean squared error (RMSE) and coverage of the 95% credible interval (CIC-95) for each of the evaluation timepoints. We report the average statistics across the evaluation points for each case. The evaluation is repeated for the common and individual-specific HPs models.

## 3 Results
### Data
Longitudinal blood test data was available for 149 data points (114 for training, 35 for testing) across the 31 individuals. All patients had received photon radiotherapy for a brain tumour of mixed diagnosis. Patients were of median age 21.0 (3.79 – 25.5) years at radiotherapy. IGF-1 measurements were recorded at a median (range) of 4.82 (0 - 14.4) years after radiotherapy. All patients were on growth hormone replacement therapy and had begun this intervention treatment 7.19 (0.49 – 14.6) years after radiotherapy which should return IGF-1 to normative ranges. In terms of data availability, the median (range) number of recordings of IGF-1 for each patient was 4 (1-16). Six patients had only one recording and hence were included in the training dataset, and the rest of the data was randomly selected for training and testing.

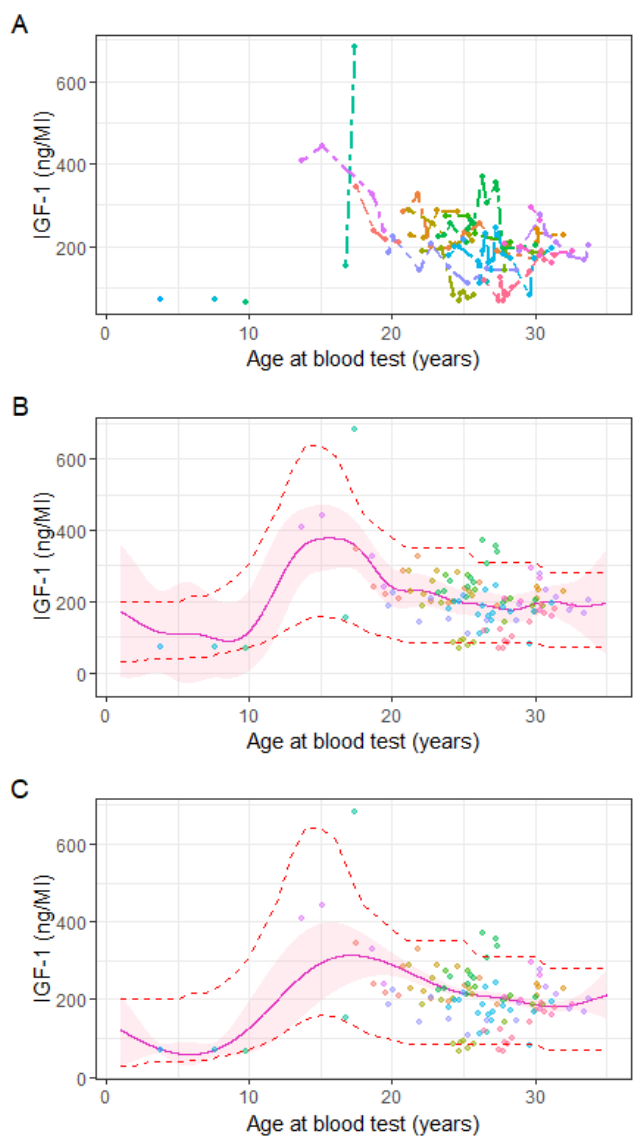

**Figure 1.** A) Individual trajectories for patients within the training dataset. Each colour represents a patient, with lines connecting time-points from the same individual. B) Fitting of a GP model with common hyperparameters, the light pink represents the 95% credible interval. C) Fitting GP model without individual-specific hyperparameters. The red dashed lines display normative trends for IGF-1.

### Training
In Figure 1A an individual's 1GF-1 trajectory is mapped out over the age at which the test was taken. Figure 1B shows the GP model assuming common mean hyperparameters. The maximum log-likelihood for this model fit was -804.56 with a minimum of -821.74 across the 25 training runs. In the GP model with individual-specific HPs (Figure 1C) the maximum log-likelihood was -744.88 within a minimum of -779.31. In comparison to normative reference values (red dashed lines on Figure 1 [8]), when there is a common HP most timepoints are within the ranges until very few data-points for training are available (i.e., <12 years old) and the GP follows the shape of the normative trajectory. For individual-specific HP, the credible interval is within the range for all timepoints, but the shape is smoother with a smaller peak. The models trained in 64 and 76 seconds respectively.

**Testing**

The results of the 8 test cases for individual-specific GPs are reported in Table 1.

| Case | Prediction | Evaluation | RMSE (common HP) | RMSE (individual-specific HP) |
|---|---|---|---|---|
| 1 | 2 | 3 | 24.96 | 31.94 |
| 2 | 2 | 2 | 38.24 | 92.55 |
| 3 | 2 | 3 | 25.56 | 41.24 |
| 4 | 1 | 1 | 10.12 | 18.74 |
| 5 | 4 | 5 | 62.29 | 66.06 |
| 6 | 1 | 1 | 46.07 | 0.02 |
| 7 | 3 | 3 | 10.89 | 22.78 |
| 8 | 1 | 1 | 56.65 | 5.53 |
|   |   |   | *31.90* | *27.36* |

**Table 1.** A table of the test results for the GP models with common and individual-specific HPs. The case is an individual patient, prediction gives the number of timepoints included for prediction, and evaluation is the number of timepoints included for evaluation. The RMSE represents the average across the evaluation timepoints (ideal value is 0), and overall means are included in the bottom row.

On average, the lowest RMSE is observed in the individual-specific HP model, however, not all evaluation points were covered by the 95% credible interval suggesting a slight underestimation of the uncertainty. There was 100% CIC-95 in all cases apart from case 5 who had 50% coverage. For the common HP setting all evaluation points had 100% CIC-95. Considering individuals, in 6 out of 8 cases, the common HP approach has an improved model fit for prediction. But the overall average RMSE is skewed by the predictions where only one timepoint is available. For cases with one timepoint it appears there is an advantage to the more flexible model.

Model fit examples are displayed in Figure 2. In case one (A and B) the datapoints show fluctuations around the common mean, so, in both cases the error is small. In case 5 (C and D) an uncharacteristic peak is observed in the years post-treatment, and here the population mean process does not fit well in either case. In case 6 and 8 only one timepoint is available for fitting the individual prediction (and one for testing), so the individual-specific HP performs better (E-F and G-H respectively).

## 4 Discussion

We have studied the use of MAGMA to perform Gaussian Processes modelling on the dependency between IGF-1 and age in a dataset of children treated with radiotherapy for brain tumours and under growth hormone treatment. We have shown the GPs can correctly model the underlying expected trends from normative samples.

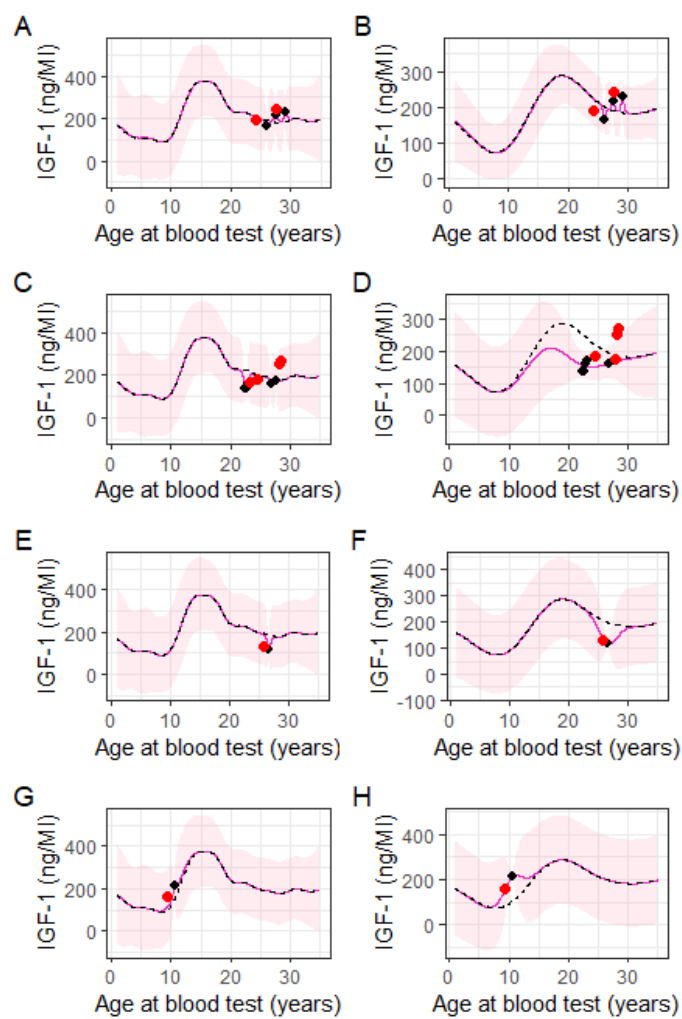

**Figure 2.** Examples of individual GP models (pink line) for 4 out of 8 example cases using the common mean process (dotted line) trained using common HPs (left) and individual-specific HPs (right). The prediction points for that case are in black, and evaluation points (unseen in training and prediction) are in red.

Individual predictions were performed with an ~30ng/ml root mean squared error, and in all but one case (94%) evaluation points were covered by the 95% credible interval. In future work, we aim to incorporate the longitudinal predictions modelled by GPs into our existing image-based data-mining pipeline [9].

Gaussian processes have previously been explored in radiotherapy for motion-modelling for image-guided treatment [11] and motion compensation in robotics treatment [12]. However, as far as the authors are aware they have not yet been utilized for dose-response outcome modelling.

The large credible intervals displayed by the model are unsurprising given the noise in the dataset. It is known that IGF-1 scores are also influenced by the time after radiotherapy, the age at radiotherapy, puberty status, weight, diet, and day-to-day variations in recording [13,14]. In addition, we are typically interested in measuring what is 'abnormal' so the ability to model normative ranges may appear a redundant test. However, all these patients were administered growth hormone therapy at some point post-radiotherapy. The potential confounding variables and external factors influencing this data raises concern for its use for showcasing the advantages of GPs in dose-response modelling. Incorporating extra covariates including time-varying ones, using MAGMA will be explored to overcome some of the issues presented here for future work with this cohort.

## 5 Conclusion

We demonstrated how multi-task GPs with a common mean process (MAGMA) can be used in routine and 'inadequately sampled' outcome data. The model was able to identify population-level trends that match known values and make individual predictions on 'missing' data-points where the true values were within the 95% credible interval for 94% of cases. This tool will aid the analysis of late effects of radiotherapy in childhood cancer as we could interpolate missing timepoints in routine follow-up data, and eventually study patterns of development across different treatment groups or dose levels. This will also complement our ongoing image-based data mining work for endocrine outcomes.